\documentclass[oldversion,letter]{aa}  
\usepackage{graphicx}
\usepackage{txfonts}
\usepackage{natbib}
\bibpunct{(}{)}{;}{a}{}{,}

\begin{document}

\title{The coronal Ne/O abundance of $\alpha$~Centauri\thanks{Based on observations obtained with \emph{XMM-Newton}, an ESA science mission with instruments and contributions directly funded by ESA Member States and NASA}}

\author{C. Liefke \and J. H. M. M. Schmitt}

\offprints{C. Liefke,\\ \email{cliefke@hs.uni-hamburg.de}}

\institute{Hamburger Sternwarte, Universit\"at Hamburg, Gojenbergsweg 112,
D-21029 Hamburg, Germany}

\date{Received 9 August 2006 / Accepted 20 August 2006}

\abstract{Recent improvements in the modeling of solar convection and line formation led to downward revisions of the solar photospheric abundances of the lighter elements, which in turn led to changes in the radiative opacity of the solar interior and hence to conflicts with the solar convection zone depth as inferred from helioseismic oscillation frequencies. An increase of the solar Ne/O abundance to values as observed for nearby stars has been proposed as a solution.  Because of the absence of strong neon lines in the optical, neon abundances are difficult to measure and the correct solar and stellar Ne/O abundances are currently hotly debated.
Based on X-ray spectra obtained with \emph{XMM-Newton}, we determine a reference value of Ne/O for the inactive, solar-like star $\alpha$~Cen (primarily $\alpha$~Cen~B, which is the dominant component in X-rays), with three independent, line-based methods, using differential emission measure reconstruction and an emission measure-independent method. 
Our results indicate a value of $\approx 0.28$ for $A_{Ne}$/$A_O$ in $\alpha$~Cen, approximately twice the value measured for the Sun, but still below the average value obtained for other stars. The low Ne/O abundance of the Sun is peculiar when compared to $\alpha$ Cen and other stars; our results emphasize the necessity to obtain more and accurate Ne/O abundance measurements of low activity stars.}

\keywords{stars: abundances -- stars: activity -- stars: coronae -- X-rays: stars}

\maketitle

\section{Introduction}

The canonical values of the solar photospheric elemental abundances have recently become far less canonical.  Starting from the work of \citet{Anders_Grevesse}, a standard reference for many years, the absolute abundance values of the more abundant trace elements like carbon, nitrogen, oxygen, neon, or iron have been significantly reduced over the past decade, initially by \citet{Grevesse_Sauval}, and again more recently \citep[and references therein]{Asplund}. 
These lighter elements considerably influence the physics of the solar interior since they provide a substantial contribution to its radiative opacity.  A change in the elemental abundances usually changes the depth of the solar convection zone,
which can be inferred from the measured helioseismological oscillation frequencies.  With the "old" abundances by \citeauthor{Grevesse_Sauval}, good agreement could be found between the standard solar model of the appropriate age and the observed oscillation spectrum, while the "new" but rather controversial abundances proposed by \citeauthor{Asplund} and collaborators turned out to be inconsistent with helioseismology \citep{Bahcall_helioseismology}. 

In order to rescue the agreement between the standard solar model and helioseismology
the opacity reduction by the downward revision of the CNO abundances must
be sufficiently compensated by increased abundances of other elements. The only suitable element is neon, since its (photospheric) abundance is not well determined due to the absence of strong photospheric lines; rather, the solar neon abundance is obtained either from solar energetic particles or from coronal measurements in the X-ray or EUV bands. The solar neon abundance is usually quoted relative to oxygen, and the solar Ne/O abundance ratio has remained more or less constant during the course of the revisions of the solar photospheric abundances, with values ranging from
0.14 from the compilation of \citeauthor{Anders_Grevesse}, 0.18 from \citeauthor{Grevesse_Sauval}, and 0.15 for the new set from \citeauthor{Asplund}. An increase of the solar Ne/O abundance by a factor of 2.5--3.5 \citep{Antia_Basu,Bahcall} would provide sufficient opacity to reconcile the low oxygen abundance with helioseismology.
Evidence for an increased neon abundance has been proposed by \citet{Drake_Testa} in their survey of the coronal Ne/O abundance in a sample of nearby stars, finding an average value of $A_{Ne}/A_O = 0.41$.

On the other hand, a
re-analysis of \emph{SOHO} CDS spectra and re-investigation of archival solar coronal X-ray spectra confirm the long-established, "canonical"  lower $A_{Ne}/A_O$ values \citep{Young_neon_to_oxygen,Schmelz_neon_to_oxygen}. Also, a closer look at the sample of stars used by \citeauthor{Drake_Testa} reveals that most of these stars are RS~CVn systems or  well-known young and active stars, known to show the inverse FIP effect \citep{Brinkman_HR_1099}, i.\,e., an enhancement of elements with high first ionization potential. Since neon is (apart from helium) the element with the highest FIP and the occurrence of the inverse FIP effect is related to activity \citep[and references therein]{Guedel_review}, the sample may be biased to higher neon abundances and not
be representative of the "true" cosmic neon abundance.

In order to settle the issue of a possible bias, a comparison to exclusively low-activity solar-like stars is needed. Due to their low X-ray luminosity, high-resolution X-ray spectra with reasonable signal-to-noise ratio of such inactive stars can be obtained only for very few objects, like $\epsilon$~Eri, Procyon, or $\alpha$~Cen. Additionally, their low coronal temperatures complicate the measurement of the otherwise prominent \ion{Ne}{ix} and \ion{Ne}{x} He-like and H-like lines that have peak formation temperatures of $\log T = 6.6$ and $\log T = 6.75$ respectively; for example, the
\emph{Chandra} LETGS spectra of $\alpha$ Cen A and B presented by \citet{Raassen}) do not show the
\ion{Ne}{x} Ly~$\alpha$ line.  Our new \emph{XMM} RGS spectra of $\alpha$~Cen provide good sensitivity and signal-to-noise to detect and measure the relevant H- and He-like lines of O and Ne to allow an accurate determination of its neon-to-oxygen abundance.

\section{Observations and data analysis}

$\alpha$ Centauri is target of an \emph{XMM-Newton}
monitoring campaign of its long-term X-ray behavior.  Since March 2003 X-ray
observations have been performed regularly at intervals of approximately six months, lasting between 5 and 9~ks each. Results from the first five observations of the program focusing on variability and possible activity cycles of both components have been presented by \citet{Robrade_alpha_Cen}. Two additional datasets (ObsIDs 0143630201 and 0202611201) are now available, resulting in 52~ks of accumulated observing time. All seven datasets were reduced with the \emph{XMM-Newton} Science Analysis System (SAS) software, version 7.0, making use of standard selection and filtering criteria. 

\begin{figure}
\resizebox{\hsize}{!}{\includegraphics{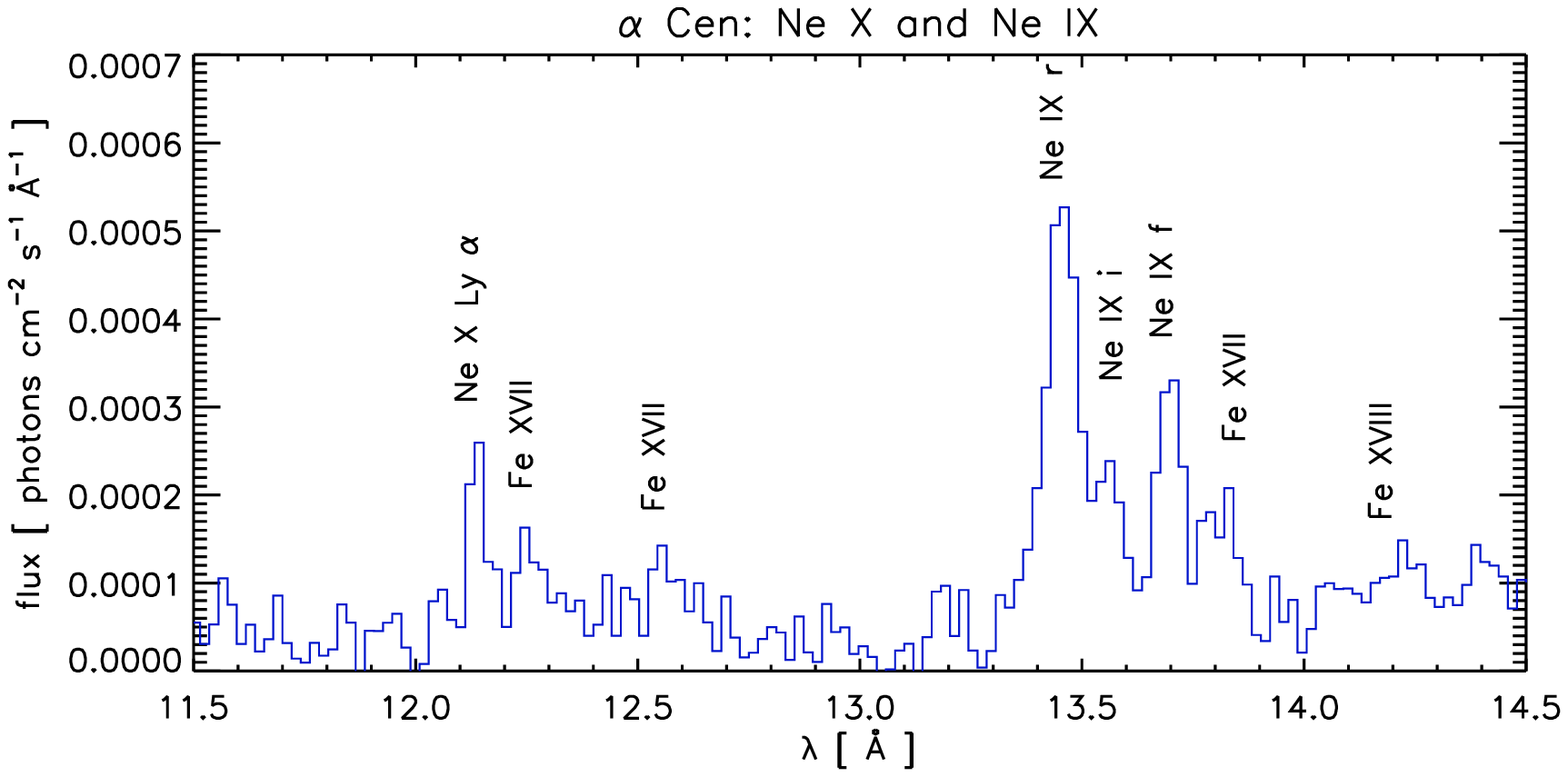}} 
\resizebox{\hsize}{!}{\includegraphics{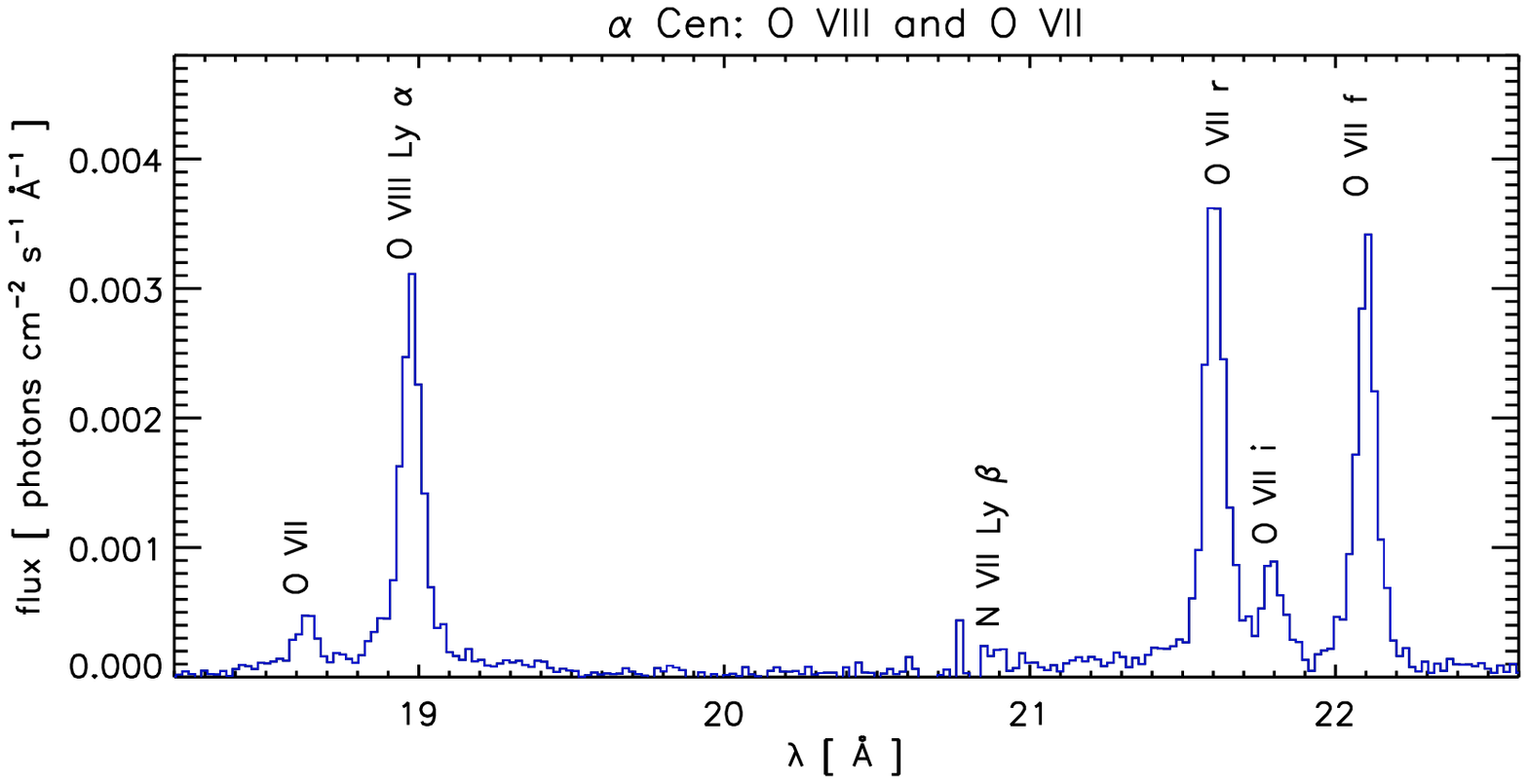}}
\caption[]{\label{Ne_O_spec} Segments of the merged RGS spectrum of $\alpha$~Cen, showing the spectral regions covering the neon (top) and oxygen (bottom) Ly~$\alpha$ lines and He-like triplets. The spectra were created with the SAS task {\tt rgsfluxer} from RGS~1 and 2 in 1st and 2nd order.}
\end{figure}

\citeauthor{Robrade_alpha_Cen} showed that the K0\,V star $\alpha$~Cen~B is X-ray brighter than the solar twin $\alpha$~Cen~A (spectral type G2\,V) by factors ranging from 3.6 to 75; this applies also to the two latest observations.  A spatially resolved spectral analysis of the $\alpha$~Cen~system is not possible with \emph{XMM-Newton}; the measured total X-ray flux refers essentially to $\alpha$~Cen~B ($\approx90$--95\%). The following analysis is based solely on the RGS data.  
We use the SAS task {\tt rgscombine} to merge the individual RGS exposures to co-added spectra with corresponding response matrices. The resulting combined spectra thus constitute a mixture of $\alpha$~Cen~A and B (with B by far dominating) and an average over the quiescent and flaring states of $\alpha$~Cen~B and the long-term variability of $\alpha$~Cen~A \citep{Robrade_alpha_Cen}. This is not critical for our purposes since we focus on the coronal abundances, which we do not expect to change during $\alpha$~Cen's different states of activity; also, \citet{Raassen} showed $\alpha$~Cen~A and B to have similar abundances, thus we do not anticipate effects from the changing contributions of the two components. In Fig.~\ref{Ne_O_spec} we plot the relevant portions of the RGS spectrum of $\alpha$~Cen that cover the He-like and H-like lines of neon and oxygen. While the \ion{O}{vii} and \ion{O}{viii} lines are recorded with a very good signal-to-noise ratio, the signal is much lower for \ion{Ne}{ix} and \ion{Ne}{x}, but the lines are still easily detectable and the \ion{Ne}{ix} resonance, intercombination and forbidden lines can be resolved.
Using the CORA program \citep{CORA} we measured individual line fluxes in the 
RGS~1 and 2 spectra in 1st and 2nd order assuming Lorentzian line profiles. Error-weighted means were calculated for further analysis. 

\section{Abundance determination}
\subsection{Differential emission measure modeling}
\begin{figure}
\resizebox{\hsize}{!}{\includegraphics{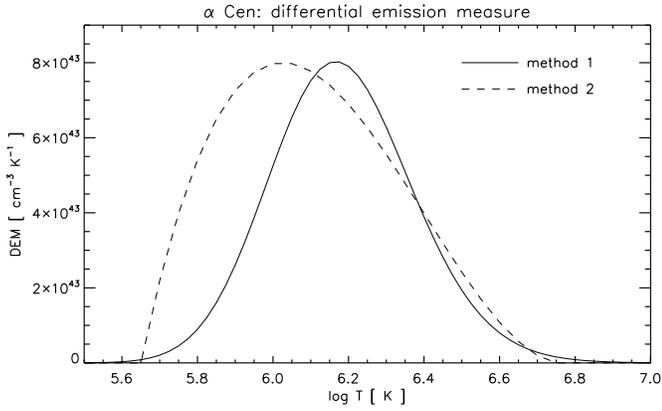}}
\caption[]{\label{alpha_Cen_DEM} $DEM$ distributions obtained by modeling a 3$^{rd}$ order polynomial to $\log DEM(\log T)$ (method~1) and a 4$^{th}$ order polynomial with two zeros to $DEM(\log T)$ (method~2). Note that the shape of the $DEM$ is well-determined only for $\log T > 6.0$.}
\end{figure}

From our line flux measurements we proceed to determine the coronal Ne/O ratio of $\alpha$~Cen
using three different methods.
We reconstructed the differential emission measure ($DEM$) from abundance-independent ratios of the H-like Ly~$\alpha$ and the He-like resonance lines from N, O, and Ne, analogous ratios of H-like Ly~$\alpha$ and the lines originating from the $1s3p - 1s^2$ transition (''He-like Ly~$\beta$'') of C and O and the weakly temperature-dependent ratio of the \ion{Fe}{xvii} lines at 15.01~\AA\ and 16.78~\AA.  
In addition we used continuum flux measurements at wavelengths around 20~\AA\ where the spectrum is essentially line-free for normalization. Our $DEM$ reconstruction method is similar to the one applied by \citet{Algol_EMD} and makes use of CHIANTI~5.2 line and continuum emissivities \citep{Chianti7}. 
In a first approach, $\log DEM$ was modelled as a function of $\log T$ using polynomials of different orders without further constraints; the best-fit is obtained with a 
3$^{rd}$ order polynomial ($\chi^2_{red} = 0.6$). However, the available line ratios cover only temperatures $\log T > 6.0$ and abundance-independent line ratios with suitable signal-to-noise for lower temperatures are not available in the RGS spectral range.
In a second approach, we model the linear $DEM$ again with polynomials as a function of $\log T$.  Additionally the $DEM$ was forced to have two zeros defining the boundaries of the coronal $DEM$ distribution. Here, the best fit is obtained with a 4$^{th}$ order polynomial ($\chi^2_{red} = 2.2$).  The resulting $DEM$ distributions are shown in
Fig.~\ref{alpha_Cen_DEM}; they agree at higher temperatures, but differ significantly for $\log T < 6.2$, indicating the uncertainties due to the poor coverage of lower temperatures. The formation temperatures of the neon and oxygen He-like and H-like lines are however well-determined and there the $DEM$ distributions look quite similar. In Table~\ref{results} we compare the observed line ratios with
the line ratios "predicted" by the two methods; the two methods agree quite well except for the \ion{Ne}{x}~Ly~$\alpha$~/~\ion{Ne}{x}~r ratio, which is reproduced much better
by method~1.

By forcing the $DEM$ distributions obtained with both methods and the corresponding line contribution functions to reproduce the measured line fluxes, we determine the absolute 
(and relative) abundances of neon and oxygen. The results are listed in Table~\ref{abundances}; the relative neon-to-oxygen abundance with method 1 is $A_{Ne}$/$A_O = 0.27 \pm 0.03$, while method 2 yields  $A_{Ne}$/$A_O = 0.31 \pm 0.08$. Note that errors are based on count statistics alone, i.\,e. the smaller error for the first approach results in the  better quality of the fit, giving consistent individual abundances for the two neon lines, while they clearly deviate in the second approach. 

\subsection{Emission measure-independent linear combinations of line fluxes}
\begin{figure}
\resizebox{\hsize}{!}{\includegraphics{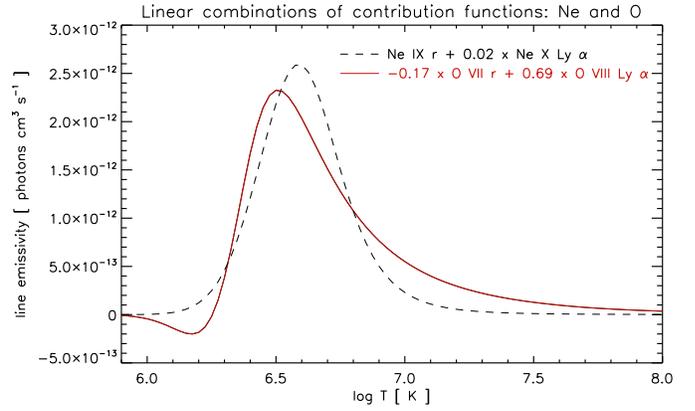}}
\caption[]{\label{Ne_to_O} Linear combinations of contribution functions of H-like Ly~$\alpha$ and He-like resonance lines of neon and oxygen.}
\end{figure}

\citet{Acton_neon_to_oxygen} proposed to determine the solar coronal Ne/O abundance ratio from the ratio of the measured line fluxes of the \ion{Ne}{ix} resonance and \ion{O}{viii} Ly $\alpha$ lines since their contribution functions have similar peak formation temperatures and a similar temperature dependence. This approach thus avoids uncertainties in the abundance determination introduced by the initially unknown underlying temperature structure of the emitting plasma.  
Acton's method has been refined by \citet{Drake_Testa} by adding a portion of the \ion{Ne}{x} Ly~$\alpha$ flux 
to reduce the temperature residuals. We further refined this approach by also taking the \ion{O}{vii} resonance line into account and calculating optimal linear combinations of the measured fluxes. The coefficients for the linear combinations are obtained from a minimization procedure incorporating the corresponding ratios of the theoretical emissivities of the involved lines from the CHIANTI database. Relative to the \ion{Ne}{ix}~r line, we obtain scaling factors of 0.02, $-$0.17, and 0.69 for \ion{Ne}{x} Ly~$\alpha$, \ion{O}{vii}~r, and \ion{O}{viii} Ly~$\alpha$ respectively, for the line fluxes and line contribution functions in photon (not energy) units, see also Fig~\ref{Ne_to_O}. These coefficients give $A_{Ne}$/$A_O = 0.28 \pm 0.05$, in very good agreement with the value obtained with the $DEM$ reconstruction methods.

\setlength{\tabcolsep}{5pt}
\begin{table}
\begin{center}
\begin{tabular}{lccc}
\hline\hline
line ratio &measured&method 1	&method 2\\
\hline
\ion{N}{vii} Ly $\alpha$ / \ion{N}{vi} r    &   1.40 $\pm$ 0.16 &   1.36 & 1.30\\
\ion{O}{viii} Ly $\alpha$ / \ion{O}{vii} r  &   0.79 $\pm$ 0.04 &   0.79 & 0.83\\
\ion{Ne}{x} Ly $\alpha$ / \ion{Ne}{ix} r    &   0.48 $\pm$ 0.10 &   0.47 & 0.28\\
\ion{C}{vi} Ly $\alpha$ / \ion{C}{v} $\beta$&  24.17 $\pm$ 10.19&  25.08 & 19.97\\
\ion{O}{viii} Ly $\alpha$ / \ion{O}{vii} $\beta$  &   7.64 $\pm$ 0.67 &   7.85 & 8.22\\
\ion{Fe}{xvii} 15.01\,\AA\ / 16.78\,\AA\    &   1.70 $\pm$ 0.20 &   1.45 & 1.42\\
\hline
\end{tabular}
\caption[]{\label{results} Line ratios (photon fluxes) used in the fitting procedure.}
\end{center}
\end{table}
\setlength{\tabcolsep}{6pt}

\begin{table}
\begin{center}
\begin{tabular}{lcccc}
\hline\hline
		&\multicolumn{2}{c}{$DEM$ modeling}& linear&Asplund\\
		&method 1	&method 2 & combinations &et al.\\
\hline
Ne		& $7.95\pm0.04$ & $8.01\pm0.13$ &---&7.84\\
O		& $8.52\pm0.01$ & $8.51\pm0.01$ &---&8.66\\
$A_{Ne}$/$A_O$	& $0.27\pm0.03$ & $0.31\pm0.08$ & $0.28\pm0.05$& 0.15\\
\hline
\end{tabular}
\caption[]{\label{abundances} Absolute abundances of neon and oxygen and the Ne/O abundance ratio of $\alpha$~Cen obtained with different methods. }
\end{center}
\end{table}

\section{Results and discussion}

The results of our abundance modeling (cf., Table~\ref{abundances}) are very
robust and yield values of $A_{Ne}$/$A_O \approx 0.28$, independent of the
applied method.  This value is twice as large as the "canonical" $A_{Ne}$/$A_O$
for the solar corona.  $\alpha$~Cen is probably the most suitable star
for a comparison with the Sun avoiding a possible FIP/I-FIP bias. Our $DEM$ of $\alpha$~Cen resembles that of the quiet Sun \citep[e.\,g.][]{Brosius,Landi_Landini_DEM}, which typically peaks around $\log T \approx 6.0$--6.2. 
However our values refer to  $\alpha$~Cen~B,
which is by far the more active component known to show flaring activity \citep{NEXXUS,Robrade_alpha_Cen}.

Separate X-ray spectra of $\alpha$~Cen~A and B are available from 
a 79~ks \emph{Chandra} LETGS exposure. However, both spectra have extremely
low signal in the wavelength range covering the \ion{Ne}{ix} and \ion{Ne}{x} lines. Values of $A_{Ne}$/$A_O = 0.18 \pm 0.07$ and $0.24 \pm 0.09$ for $\alpha$~Cen~A and B respectively were derived from global fitting by \citet{Raassen} and are thus based primarily on \ion{Ne}{vii} and \ion{Ne}{viii} lines located at longer wavelengths. Many of these lines suffer from significant blending as well as
low signal and their atomic physics parameters should be considered as more uncertain than those of the H-like and He-like lines.  Formally, the Ne/O
abundances of $\alpha$~Cen~A and B are consistent with each other, and the
value for the B~component is consistent with our XMM-Newton result. 

Measurements of the solar Ne/O abundance ratio tend to show a broad scatter (cf. the compilation provided by \citet{Drake_Testa} in the supplementary information, with values of $A_{Ne}$/$A_O$ ranging from 0.08 to 0.47), but the majority of them, based on miscellaneous data like solar energetic particles, X-ray or EUV spectra, are in good agreement with the "low" Ne/O abundances.
Additionally, the most recent analyses of \citet{Young_neon_to_oxygen} and \citet{Schmelz_neon_to_oxygen}, based on the most recent atomic data, clearly support values as low as 0.15. 

All stars in the survey of \citeauthor{Drake_Testa} show higher values of $A_{Ne}$/$A_O$, incompatible with the "canonical" low solar value.  The inference of the solar Ne/O abundance from other stars is, however, problematic.
Apart from the fundamental question of why the Sun should implicitly show the same abundance pattern as other stars do (and many stellar photospheric measurements show that it does not), the most severe problem is to find truly solar-like stars,
i.\,e. relatively old and inactive single stars of similar spectral type. These conditions and the basic requirement that the stars are observable in X-rays (or in the EUV) with a sufficient signal to obtain abundance measurements of neon and oxygen, are almost mutually exclusive since stars with an X-ray luminosity as low as that of the Sun can only be observed in the very solar vicinity with today's X-ray telescopes. Instead, the typical well-studied stellar coronal X-ray source is much brighter, usually consisting of an active young late-type star or even an RS~CVn system. Such objects are not appropriate for a direct comparison with the Sun, especially if one accepts the reality of the inverse FIP effect, i.\,e. an enhancement of elements with high first ionization potential; while the physics of abundance anomalies like the inverse FIP effect and its counterpart, the FIP effect as observed on the Sun, are not fully understood, a framework to explain both effects has been provided by
\citet{Laming}.

A correlation seems to exist in the sense that the FIP effect turns into the inverse FIP effect, with the inverse FIP effect 
becoming stronger with increasing activity. Clearly, the sample used by \citeauthor{Drake_Testa} is then 
strongly biased. The only star in their sample of truly solar-like activity is Procyon, where a
value of 0.42 from \citet{coronal_photospheric} was used, which was obtained from three combined \emph{Chandra} LETGS spectra 
but still low signal at the wavelengths of the \ion{Ne}{ix} and \ion{Ne}{x} lines. For part of these data 
\citet{Raassen_Procyon} obtained $A_{Ne}$/$A_O = 0.22$ with a global fitting approach, a value also found by the
same authors from 91~ks of \emph{XMM} RGS and MOS data; thus the correct 
Ne/O abundance ratio of Procyon remains an open issue.

Moderately active K dwarfs seem to have values of $A_{Ne}$/$A_O \sim 0.37$, i.e,  slightly lower than the median 
obtained by \citeauthor{Drake_Testa}, but still at approximately twice the level of the Sun as shown by  \citet{Wood_Linsky},
who investigated Ne/O abundance ratios of $\epsilon$~Eri,  36~Oph and 70~Oph.  As pointed out above, our Ne/O value
also refers to a K star, $\alpha$~Cen~B, which is less active than any of the stars studied by \citeauthor{Wood_Linsky}.
Therefore, none of the stars studied so far has a Ne/O abundance as low as observed for the Sun.  Therefore,
the question remains why the Sun is so special. Or do we first have to find a real solar-like star?

\begin{acknowledgements}
CHIANTI is a collaborative project involving the NRL (USA), RAL (UK), MSSL (UK), the Universities of Florence (Italy) and Cambridge (UK), and George Mason University (USA).\\
CL acknowledges support from DLR under 50OR0105.
\end{acknowledgements}

\bibliographystyle{aa}
\bibliography{../literature}

\end{document}